\documentclass[letterpaper, 10 pt, conference]{ieeeconf}   
\IEEEoverridecommandlockouts                               
\usepackage{optidef}

\usepackage{amssymb,amsmath,color}
\usepackage{graphicx}
\usepackage{xspace,stackrel,hyperref}
\usepackage{mathtools}
\usepackage{tikz,pgfplots}
\usepackage{etoolbox}
\usepackage{autonum}
\usepackage{dsfont}
\usepackage{etoolbox}

\usepackage{multirow}
\usepackage{cite}

\usepackage{arydshln}

\usepackage{esvect}
\usepackage{optidef}
\usepackage{multirow}

\usepackage[ruled,vlined]{algorithm2e}

\definecolor{mycolor4}{RGB}{230,97,1}
\definecolor{mycolor2}{RGB}{178,171,210}
\definecolor{mycolor3}{RGB}{253,184,99}
\definecolor{mycolor1}{RGB}{94,60,153}

\makeatletter 
\pretocmd\@bibitem{\color{black}\csname keycolor#1\endcsname}{}{\fail}
\newcommand\citecolor[1]{\@namedef{keycolor#1}{\color{blue}}}
\makeatother

\DeclareMathAlphabet{\mathcal}{OMS}{cmsy}{m}{n}

\def\beq{\begin{equation}}
\def\eeq{\end{equation}}

\newcommand{\mc}{\mathcal}

\newcommand{\Z}{\mathbb{Z}}

\newcommand{\R}{\mathds{R}}
\newcommand{\N}{\mathbb{N}}

\newcommand{\defineas}{\coloneqq}

\newcommand{\norm}[1]{\left\lVert#1\right\rVert}

\definecolor{mycolor1}{RGB}{230,97,1}
\definecolor{mycolor2}{RGB}{178,171,210}
\definecolor{mycolor3}{RGB}{253,184,99}
\definecolor{mycolor4}{RGB}{94,60,153}
\definecolor{mycolor5}{rgb}{0,0,0}

\tikzset{
  pics/car/.style args={#1}{
     code={
     \begin{scope}[scale=0.15]
      \shade[top color=#1, bottom color=white, shading angle={135}]
        [draw=black,fill=red!20,rounded corners=0.2ex] (1.5,.5) -- ++(0,1) -- ++(1,0.3) --  ++(3,0) -- ++(1,0) -- ++(0,-1.3) -- (1.5,.5) -- cycle;
    \draw[ rounded corners=0.5ex,fill=black!20!blue!20!white]  (2.5,1.8) -- ++(1,0.7) -- ++(1.6,0) -- ++(0.6,-0.7) -- (2.5,1.8);
    \draw[thick]  (4.2,1.8) -- (4.2,2.5);
    \draw[draw=black,fill=gray!50,thick] (2.75,.5) circle (.5);
    \draw[draw=black,fill=gray!50,thick] (5.5,.5) circle (.5);
    \end{scope}
     }
  }
}

\newtheorem{assumption}{Assumption}

\newtheorem{theorem}{Theorem}
\newtheorem{proposition}{Proposition}
\newtheorem{corollary}{Corollary}
\newtheorem{definition}{Definition}

\pgfplotsset{compat=1.17} 
\usetikzlibrary{pgfplots.groupplots}

\newtoggle{full_version}
\toggletrue{full_version}

\title{\LARGE \bf
On Finding the Leader's Strategy in Quadratic Aggregative Stackelberg Pricing Games
}

\author{Marko Maljkovic, Gustav Nilsson, and Nikolas Geroliminis
\thanks{M.~Maljkovic, G.~Nilsson, and N.~Geroliminis are with the School of Architecture, Civil and Environmental Engineering, École Polytechnique Fédérale de Lausanne (EPFL), 1015 Lausanne, Switzerland. {\tt\small \{marko.maljkovic, gustav.nilsson, nikolas.geroliminis\}@epfl.ch}.}%
\thanks{This work was supported by the Swiss National Science Foundation under NCCR Automation, grant agreement 51NF40\_180545.}
\iftoggle{full_version}{}{\thanks{An extended version containing all the proofs is available at \url{http://arxiv.org/abs/2203.09327}}}%
}

\begin{document}

\maketitle
\thispagestyle{empty}
\pagestyle{empty}

\begin{abstract}
This paper analyzes a class of Stackelberg games where different actors compete for shared resources and a central authority tries to balance the demand through a pricing mechanism. Situations like this can for instance occur when fleet owners of electric taxi services compete about charging spots. In this paper, we model the competition between the followers as an aggregative game, i.e., each player's decision only depends on the aggregate strategy of the others. While it has previously been shown that there exist dynamic pricing strategies to achieve the central authority's objective, we in this paper present a method to compute optimal static prices. Proof of convergence of the method is presented, together with a numerical study showcasing the benefits and the speed of convergence of the proposed method. 

\end{abstract}

\section{Introduction}
The inherent leader-follower structure of the Stackelberg games found its use in many real-world applications as the general framework allows the users to describe the interactions between different agents on multiple levels of hierarchy. As such, the Stackelberg games with a single leader and multiple followers have provided a suitable model for different problems within the realms of energy management~\cite{ex1,AUSSEL2020299}, operational optimization~\cite{ex2,ex3} and transportation~\cite{Groot2017HierarchicalGT, Hierarchical}. In this paper, we are interested in the scenarios in which different agents compete about shared resources in a setting where a central authority attempts to balance the demand through the pricing of the resources. More specifically, we are motivated by a problem that tackles a bi-level hierarchical setup in the domain of smart mobility, previously analyzed in~\cite{ecc2022,Hierarchical}. On the lower level, we look at the competition between the ride-hailing company operators interested in minimizing the personal operational cost that, among others, includes the charging of their electric fleets at publicly available stations. On the higher level, we consider the interaction between the central authority in charge of determining the prices of charging and the ride-hailing market. Compared to~\cite{ecc2022, Hierarchical}, where dynamic, demand-based pricing policies were designed for each ride-hailing company, in this paper, we focus on computing a static pricing vector that is unique for all the companies and hence yields a more realistic setting. Inspired by this particular game structure, we generalize the problem into a class of aggregative Stackelberg games that we refer to as the ``Stackelberg pricing games''. For this class of problems, to compute the local Stackelberg equilibrium originally defined in~\cite{9424958}, we design a gradient-based, iterative procedure. 
\begin{figure}[t]
    \centering
    \resizebox{0.35\textwidth}{!}{
    \begin{tikzpicture}[scale=0.7]
        \shadedraw[top color= mycolor1, bottom color=white, draw=mycolor1] (-1.25, -0.5) rectangle (1.25,0.5);
        \node (c2m) at (0.0, 0.0){\footnotesize$L$};
        
        \draw[] (-1.25, -3.0) rectangle (1.25,-2.0);
        \node[align=center] (cag) at (0,-2.5) {\footnotesize Central \\ \footnotesize aggregator};

        \shadedraw[top color= mycolor2, bottom color=white, draw=mycolor2] (-2.5, -4.5) rectangle (0.0, -5.5);
        \node (f2) at (-1.25,-5.0){\footnotesize$F_2$};
        
        \shadedraw[top color= mycolor2, bottom color=white, draw=mycolor2] (-5.1, -4.5) rectangle (-2.6, -5.5);
        \node (f1) at (-3.85,-5.0){\footnotesize$F_1$};
        
        \shadedraw[top color= mycolor2, bottom color=white, draw=mycolor2] (2.6, -4.5) rectangle (5.1, -5.5);
        \node (fN) at (3.85,-5.0){\footnotesize$F_{\left|\mc I\right|}$};
        
        \node (d1) at (1.3, -5.0)[circle,fill,inner sep=0.75pt]{};
        \node (d2) at (1.5, -5.0)[circle,fill,inner sep=0.75pt]{};  \node (d3) at (1.1, -5.0)[circle,fill,inner sep=0.75pt]{};   
        
        \draw[dashed] (-5.2, -4.3) rectangle (5.2, -5.7);
        
        \draw[dashed] (-5.4, 0.7) rectangle (5.4, -5.9);
        
        \draw[->] (-3.85 , -4.5) -- (-0.75,-3);
        \draw[->] (-1.25 , -4.5) -- (-0.25,-3);
        \draw[->] (3.85  , -4.5) -- (0.75,-3);
        
        \draw[->] (0.0, -3) -- (0.0, -4.3);
        
        \node[] (D1) at (-3.0, -3.85){\footnotesize$\textbf{D}_{\pi_t}x^{1*}$};
        \node[] (D2) at (-1.4, -3.85){\footnotesize$\textbf{D}_{\pi_t}x^{2*}$};
        \node[] (D3) at (0.3, -3.85){\footnotesize$\pi_t$};
        \node[] (D4) at (3.65, -3.85){\footnotesize$\textbf{D}_{\pi_t}x^{\left|\mc I\right|*}$}; 
        
        \draw[-] (1.25, -2.5) -- (1.75, -2.5);
        \draw[-] (1.75, -2.5) -- (1.75, 0.0);
        \draw[->] (1.75, 0.0) -- (1.25,0.0);
        
        \draw[-] (-1.25,0.0) -- (-1.75, 0.0);
        \draw[-] (-1.75, 0.0) -- (-1.75, -2.5);
        \draw[->] (-1.75, -2.5) -- (-1.25, -2.5);
        
        \node[] (Dt) at (2.35, -1.25){\footnotesize$\textbf{D}_{\pi_t}x^{*}$};       
        \node[] (pi) at (-2.15, -1.25){\footnotesize$\pi_{t+1}$};
        
        \node[] (G0) at (-4.6, -4.0){\footnotesize$G_0\left(\pi_t\right)$};
        \node[] (G1) at (-3.5, 1.0){\footnotesize Stackelberg game $G_1$};
        
    \end{tikzpicture}}

    \caption{Schematic sketch of the problem setting. The $\left|\mc I\right|$ followers communicate with the leader through the central aggregator entity. It is required for computing the Nash equilibrium between the followers and is also used as a medium to collect the locally computed Jacobians $\textbf{D}_{\pi_t}x^{i*}$ in every update step of the leader's action.\vspace{-0.7cm}}
    \label{fig:problem}
\end{figure}
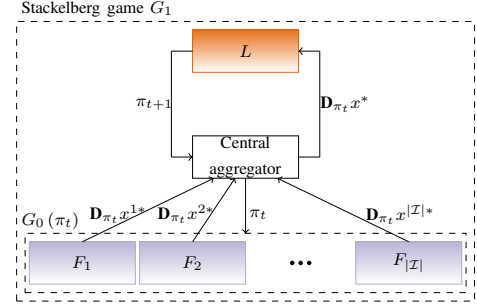
In general, finding a solution to a Stackelberg problem is challenging, and existing methods usually have to rely on specific structural assumptions of the game. In~\cite{10.2307/25614751, 918286}, the authors formulate the problem as a mathematical program with complementary constraints (MPCC)~\cite{10.2307/3690420}. In~\cite{7956147}, the authors propose an iterative method for computing the local Stackelberg equilibrium but for the case of an unconstrained game. From the methodological point of view, our approach is, to a certain extent, a combination of~\cite{9424958} and~\cite{Wang_Xu_Perrault_Reiter_Tambe_2022}. The aggregative structure of the game in~\cite{9424958} matches the one presented in~\cite{ecc2022,Hierarchical}, however, the assumed form of the leader's objective does not. Moreover, the presented method relies on a standing assumption that is in general hard to verify for the class of problems presented in~\cite{ecc2022,Hierarchical}. On the other hand,~\cite{Wang_Xu_Perrault_Reiter_Tambe_2022} elucidates the effectiveness of differentiating the KKT conditions in an attempt to estimate how the attained Nash equilibrium between the followers reacts to a change in the leader's action. Even though the problem has been tackled from the perspective of centralized, gradient-based, iterative methods, formal results on the convergence properties seem to be missing. For the particular case of games analyzed in this paper, we were able to also decentralize the calculation of Jacobians required for the computation of the local Stackelberg equilibrium, which resonates well with the privacy-preserving nature of the ride-hailing market. 

The proposed semi-decentralized approach leverages the existence of a Central aggregator typically required for decentralized computation of the Nash equilibrium of the game played between the followers~\cite{Paccagnan2019,Paccagnan2016b,Decentralized}. Inspired by~\cite{Amos2017OptNetDO,Wang_Xu_Perrault_Reiter_Tambe_2022}, we show how the followers can locally compute the Jacobians by the means of Implicit function theorem, in an attempt to estimate the influence of the leader's decision variable on the attained Nash equilibrium of the followers. By designing an iterative procedure that concerns the central aggregator and the leading player and guarantees local improvement of the leader's objective at each iteration, we can provide formal convergence guarantees. Compared to~\cite{Wang_Xu_Perrault_Reiter_Tambe_2022}, where a standing assumption was made on the conditions of the Implicit function theorem, here we explicitly address these conditions by proposing to differentiate the KKT conditions of an optimization problem that is equivalent to the standard best-response optimization problem of each follower. The complete schematic representation of the method is described in Figure~\ref{fig:problem}. We apply our method in a numerical case study identical to the one in~\cite{Hierarchical} and show how it can be used to find the local Stackelberg equilibrium corresponding to the societal optimum of the system. Finally, when it comes to fairness in pricing the companies, this method cannot experience potential problems resulting from having different prices for different companies, as was the case in~\cite{Hierarchical}.

The paper is outlined as follows: the rest of this section is devoted to introducing some basic notation. In Section~\ref{sec:model} we introduce the problem setup and the specific class of Stackelberg games that we consider. In the following section, Section~\ref{sec:alg}, we then introduce the method for finding the local Stackelberg equilibrium. Finally, we conclude the paper with Sections~\ref{sec:example} and~\ref{sec:conclusion}, where we test our method in a numerical case study and propose ideas for future research.

\textit{Notation:} Let $\R$ denote the set of real numbers, $\R_+$ the set of non-negative reals, and $\Z_+$ the set of non-negative integers. Let $\mathbf{0}_{m}$ and $\mathbf{1}_{m}$ denote the all zero and all one vectors of length $m$ respectively, and $\mathbb{I}_{m}$ the identity matrix of size $m \times m$. For a finite set $\mc A$, we let $\R_{(+)}^{\mc A}$ denote the set of (non-negative) real vectors indexed by the elements of $\mc A$ and $\left|\mc A\right|$ the cardinality of $\mc A$. Furthermore, for finite sets $\mc A$, $\mc B$ and a set of $|\mc B|$ vectors $x^i\in \R_{(+)}^{\mc A}$, we define $x \defineas \text{col}\left((x^{i})_{i\in \mc B}\right)\in \R^{|\mc A||\mc B|}$ to be their concatenation. For $A\in \R^{n\times n}$, $A\succ 0 (\succeq 0)$ is equivalent to $x^TAx>0 (\geq 0)$ for all $x\in\R^{n\times n}$. We let $A\otimes B$ denote the Kronecker product between two matrices and for a vector $x\in\R^n$, we let $\text{Dg}(x)\in\R^{n\times n}$ denote a diagonal matrix whose elements on the diagonal correspond to vector $x$. For a differentiable function $f(x):\R^n\rightarrow \R^m$, we let $\textbf{D}_xf\in\R^{m\times n}$ denote the Jacobian matrix of $f$ defined as $\left(\textbf{D}_x f\right)_{ij}\defineas\frac{\partial f_i}{\partial x_j}$. Finally, for a set-valued mapping $\mathcal{F}: \mathbb{R}^n \rightrightarrows \mathbb{R}^m, \operatorname{gph}(\mathcal{F}):=$ $\left\{(y, x) \in \mathbb{R}^n \times \mathbb{R}^m \mid x \in \mathcal{F}(y)\right\}$ denotes its graph.

\section{Problem statement}\label{sec:model}
Let us consider a Stackelberg game with a set of $N+1$ agents $\overline{\mc I}=\mc I\cup\left\{L\right\}$, where $i=L$ represents the leading agent and each $i\in\mc I$ represents one of the $N$ followers. Each player can choose a personal decision vector from a local constraint set, such that the leader first announces its strategy $\pi\in\mc P\subseteq\R^{m_L}$ to the followers and then lets the agents $i\in\mc I$ simultaneously respond with a personal decision vector $x^i\in\mc X_i\subseteq \R^{m_F}$, with $m_L,m_F\in\N$ representing the dimension of the decision space of the leader and the followers. We assume $\mc X_i$ of the form
\iftoggle{full_version}{
\begin{equation}
    \mc X_i\defineas\left\{x^i\in\R^{m_F}\mid \boldsymbol{g_1^i}\left(x^i\right)\leq\mathbf{0}_{m_i^{\text{ineq}}}\land\boldsymbol{g_2^i}\left(x^i\right)=\mathbf{0}_{m_i^{\text{eq}}}\right\}\,,
\end{equation}
}{
$\mc X_i\defineas\left\{x^i\in\R^{m_F}\mid \boldsymbol{g_1^i}\left(x^i\right)\leq\mathbf{0}_{m_i^{\text{ineq}}}\land\boldsymbol{g_2^i}\left(x^i\right)=\mathbf{0}_{m_i^{\text{eq}}}\right\}$,
}
with $m_i^{\text{ineq}},m_i^{\text{eq}}\in\N$ and the maps $\boldsymbol{g_1^i}:\mc X_i\rightarrow\R^{m_i^{\text{ineq}}}$ and $\boldsymbol{g_2^i}:\mc X_i\rightarrow\R^{m_i^{\text{eq}}}$ that describe the constraint set of each follower. If we define the sets $\mc X\defineas \prod_{i\in\mc I} \mc X_{i}$ and $\mc X_{-i}\defineas \prod_{j\in\mc I\setminus i} \mc X_{j}$, then the joint strategy of all followers can be denoted as $x \defineas \text{col}\left((x^{i})_{i\in \mc I}\right)\in\mc X$ and for every agent $i\in\mc I$ we can define $x^{-i} \defineas \text{col}\left((x^{j})_{j\in \mc I \setminus i}\right)\in\mc X_{-i}$. 

Upon the announcement of the leader's strategy $\pi\in\mc P$, the followers choose their strategies in an attempt to minimize personal objective functions $J^i(x^i, x^{-i}, \pi)$ by playing the best response to the other agents' strategies. Let $\sigma\left(x^{-i}\right) \defineas \sum_{j \in \mc I \setminus i} x^j$ be the aggregate decisions of the other players, then we assume the followers admit an aggregative game
\begin{equation}
    G_0\left(\pi\right)\defineas \left\{\min_{x^{i}\in \mc X_{i}}J^{i}\left(x^{i},\sigma\left(x^{-i}\right),\pi\right),\forall i\in \mc I\right\} \,,
    \label{eq:gameform}
\end{equation}
whose Nash equilibrium $x^{*}$ is given in the definition below:
\begin{definition}[Nash equilibrium]\label{def:NE}
    For any leader's strategy $\pi\in\mc P$, a joint strategy $x^*\in\mc X$ is a Nash equilibrium (NE) of the game $G_0$, if for all $i\in\mc I$ and all $x^i\in\mc X_i$ it holds that
    \begin{equation}
        J^i\left(x^{i*}, x^{-i*}, \pi\right)\leq J^i\left(x^{i}, x^{-i*},\pi\right) \,.
        \label{eq:NEdef}
    \end{equation}
\end{definition}
\medskip
Principally, solving a generalized Nash Equilibrium problem (GNEP) is difficult in the sense that we can rarely obtain a closed-form characterization of the full set of NE. Therefore, we focus on a subset of all NE called the variational Nash equilibria (v-NE). Based on the theory of variational inequalities~\cite{VISurvey}, if we define a map $F:\mc X\times\mc P\rightarrow\R^{Nm_F}$ 
\begin{equation}
\label{eq:PSG}
    F\left(x,\pi\right)\defineas \text{col}\left(\left(\nabla_{x^{i}}J^{i}\left(x^i, x^{-i}, \pi\right)\right)_{i\in \mc I}\right)\,,
\end{equation}
then the set of all v-NE of game~\eqref{eq:gameform} is given by
\begin{equation}
\label{eq:setNE}
    \mc N_0\left(\pi\right)\defineas\left\{x\in\mc X\left.\right|\left(y-x\right)^TF\left(x,\pi\right)\geq 0,\:\forall y\in\mc X\right\}\,.
\end{equation}
Depending on the properties of the cost functions and constraint sets of the agents, we can utilize different methods for computing the Nash equilibrium of the followers~\cite{Paccagnan2019, Paccagnan2016b, Decentralized}.
Computing the leader's optimal strategy $\pi\in\mc P$, on the other hand, requires solving a problem of minimizing the leader's personal objective of the form $J^L:\mc X\times\mc P\rightarrow\R$, where $x^{*}\in\mc N_0(\pi)$. For the solution of the Stackelberg game played between the leading agent and the $N$ followers, we adopt a definition given below.
\begin{definition}[Stackelberg game]\label{def:SG}
    Let $\overline{\mc I}$ be the set of all agents, $\hat{\pi}\in\mc P$, $\hat{x}^{*}\in\mc X$ and $\mc N_0\left(\pi\right)$ be defined in~\eqref{eq:setNE}. Then a tuple $\left(\hat{\pi}, \hat{x}^{*}\right)$ is a solution of a Stackelberg game with one leader and followers defined by $\mc I$, if it solves a bi-level optimization problem $G_1$:
    \begin{equation}
        G_1:=\left\{\begin{array}{c}
        \displaystyle\min _{\pi \in\mc P}  J^{L}\left(x^{*},\pi\right) \\
        \text { s.t. } \left(x^{*},\pi\right) \in\text{gph}\left(\mc N_0\right)\cap\left(\mc X\times\mc P\right)
        \end{array}\right\} \,.
        \label{eq:SG}
    \end{equation}
\end{definition}
\medskip
Finding the optimal leader's strategy in $G_1$ is in general difficult as it directly requires having the ability to understand how the leader's strategy influences the position of the Nash equilibrium of $G_0\left(\pi\right)$. In this paper, we are particularly inspired by the scenarios in which multiple actors compete for resources and the central authority is interested in balancing the demand by pricing the resources. In light of this system structure, we first introduce a class of games that we refer to as the ``Stackelberg pricing games''.
\subsection{Stackelberg pricing games}
In a Stackelberg pricing game, it is assumed that each follower $i\in\mc I$ has a personal objective cost $J^i(x^i, x^{-i}, \pi)$
\begin{equation}
    \label{eq:pricingcost}
    J^i\left(x^i, x^{-i}, \pi\right)\defineas \hat{J}^i\left(x^i, x^{-i}\right)+\left(x^i\right)^TS_i\pi\,,
\end{equation}
for some diagonal $S_i\succeq 0$. The first term $\hat{J}^i\left(x^i, x^{-i}\right)$ encapsulates the influence of other followers on the perceived cost whereas the second one describes the total price agent~$i$ has to pay for choosing a particular $x^i$, hence the term ``pricing game''. We assume a quadratic aggregative form of the cost term $\hat{J}^i\left(x^i, x^{-i}\right)$ given by
\begin{equation}
    \hat{J}^i\left(x^i, x^{-i}\right)=\frac{1}{2} \left(x^{i}\right)^T P_{i}x^{i}+\left(x^{i}\right)^TQ_{i}\sigma\left(x^{-i}\right)+r_{i}^{T}x^{i} \,,
\label{eq:J1igen}
\end{equation}
for some $P_i,Q_i\in\R^{m_F\times m_F}$ and $r_i\in\R^{m_F}$. Moreover, we postulate some common assumptions about the cost functions and constraint sets, as in, e.g.,~\cite{Paccagnan2016a, Paccagnan2016b, Paccagnan2019} ensuring convexity and regularity of the individual optimization problems.
\begin{assumption}[Cost functions]\label{ass:1}
For every agent $i\in\mc I$, let the cost functions of the followers be defined using~\eqref{eq:pricingcost} and~\eqref{eq:J1igen} and $P_i=P_i^T=P\succ 0$, $Q_i=Q_i^T=Q\succeq 0$, $P\succ Q$.
\end{assumption}
\medskip
\begin{assumption}[Constraint sets]\label{ass:2}
For every agent $i\in\mc I$, assume that the set $\boldsymbol{g_1^i}\left(x^i\right)\leq\mathbf{0}_{m_i^{\text{ineq}}}\cap\boldsymbol{g_2^i}\left(x^i\right)=\mathbf{0}_{m_i^{\text{eq}}}$ is nonempty, compact, satisfies Slater's constraint qualification and is separated into linear equality and inequality constraints 
\begin{equation}
    \mc X_i\defineas\left\{x^i\in\R^{m_F}\mid A_ix^i=b_i \land G_ix^i\leq h_i\right\}\,,
\end{equation}
where $\boldsymbol{g_1^i}\left(x^i\right)= G_ix^i- h_i$ and $\boldsymbol{g_2^i}\left(x^i\right)=A_ix^i-b_i$, for $A_i\in\R^{m_i^{\text{eq}}\times m_F}$, $b_i\in\R^{m_i^{\text{eq}}}$, $G_i\in\R^{m_i^{\text{ineq}}\times m_F}$ and $h_i\in\R^{m_i^{\text{ineq}}}$.
\end{assumption}
\medskip
With the pseudo-gradient mapping of $G_0\left(\pi\right)$ as in~\eqref{eq:PSG}, we can summarize the existence and uniqueness results for the Nash equilibrium of $G_0\left(\pi\right)$ in the following proposition.
\begin{proposition}\label{prop:1}
Let the game $G_0\left(\pi\right)$ be defined as in~\eqref{eq:gameform} such that for each $i\in\mc I$, the cost function $J^{i}\left(x^i, x^{-i}, \pi\right)$ is defined by~\eqref{eq:pricingcost} and~\eqref{eq:J1igen}. If Assumptions~\ref{ass:1} and~\ref{ass:2} hold, then for any leader strategy $\pi\in\mc P$, there exists a unique Nash equilibrium $x^*\in\mc X$ of $G_0\left(\pi\right)$ that can be calculated via Picard-Banach~\cite{Iterative} fixed point iteration
\begin{equation}
    x_{k+1}=\Pi_{\mc X}\left[x_{k}-\gamma F\left(x_k,\pi\right)\right] \,, 
\label{eq:iterativeproc}
\end{equation}
where $\Pi_{\mc X}$ is the projection operator and $\gamma\in\R_+$ is a step size such that the mapping $x\rightarrow x-\gamma F(x,\pi)$ is contractive in the Euclidean norm and $k$.
\end{proposition}
\iftoggle{full_version}{
\medskip
\noindent The proof of Proposition~\ref{prop:1} is given in Appendix~\ref{app:proof1}.
}{
\medskip
\noindent The proof is given in the extended version of our paper.}
\medskip

\noindent Generally speaking, under the Assumptions~\ref{ass:1} and~\ref{ass:2}, for every $i\in\mc I$, the Nash equilibrium strategy $x^{i*}\in\mc X_i$ that can be computed using~\eqref{eq:iterativeproc} is the solution of the best-response optimization problem
    \begin{equation}
    G_0^{i}\left(\pi\right):=\left\{\begin{array}{c}
    \displaystyle\min_{x^i\in\R^{m_F}}J^i\left(x^i,x^{-i*},\pi\right) \\
    \text { s.t. }\boldsymbol{g_1^i}\left(x^i\right)\leq\mathbf{0} \land \boldsymbol{g_2^i}\left(x^i\right)=\mathbf{0}
    \end{array}\right\} \,.
    \label{eq:best_resp}
\end{equation}
The optimality of $x^{i*}$ is guaranteed if and only if $x^{i*}$ solves the KKT system of equations $l_i(z_i,\pi|x^{-i*})=0$, where the mapping $l_i:\R^{n^{\text{total}}}\times\R^{m_L}\rightarrow\R^{n^{\text{total}}}$ with $n^{\text{total}}=m_F+m_i^{\text{eq}}+m_i^{\text{ineq}}$, is defined for every $z_i=(x^i,\lambda_i,\nu_i)$ as
\begin{equation}
\label{eq:KKToperator}
    l_i\left(z_i,\pi\left|x^{-i*}\right.\right)\defineas\left[\begin{array}{c}\nabla_{x^i}L_i\left(x^i,\lambda_i,\nu_i,\pi\right) \\ \operatorname{Dg}\left(\lambda_i\right)\boldsymbol{g_1^i}\left(x^i\right) \\ \boldsymbol{g_2^i}\left(x^i\right)\end{array}\right]\,,
\end{equation}
$L_i\left(x^i,\lambda_i,\nu_i\right)=J^i\left(x^{i}, x^{-i*}, \pi\right)+\lambda_i^T\boldsymbol{g_1^i}\left(x^{i}\right)+\nu_i^T\boldsymbol{g_2^i}\left(x^{i}\right)$ is the Lagrangian of the best-response problem~\eqref{eq:best_resp}, $x^{i*}$ is feasible and $\lambda_i^*\in\R_+^{m_i^{\text{ineq}}}$ and $\nu_i^*\in\R^{m_i^{\text{eq}}}$ represent the optimal dual variables associated with the inequality and equality constraints. Under Assumptions~\ref{ass:1} and~\ref{ass:2}, if some $\hat{z}_i=(\hat{x}^i,\hat{\lambda}_i,\hat{\nu}_i)$ with feasible $\hat{x}^i$ and $\hat{\lambda}_i$, satisfies $l_i\left(\hat{z}_i,\pi\left|x^{-i*}\right.\right)=0$, then $\hat{z}_i$ is the optimizer of~\eqref{eq:best_resp}.
\iftoggle{full_version}{
Note that the bi-level optimization problem of the leader is in general non-convex. Moreover, there exist subclasses of the Stackelberg pricing games for which there cannot be a unique minimizer of the leader's objective. We show an example of these games in order to elucidate the shift in our focus to computing the local optima of the leader's objective function.
\begin{corollary}\label{cor:1}
    Let the Stackelberg game $G_1$ be as in Definition~\ref{def:SG}, with the cost functions defined by~\eqref{eq:J1igen} and~\eqref{eq:pricingcost}. Let for every $i\in\mc I$, $S_i=S\succ 0$ for some diagonal $S\in\R^{m_F\times m_L}$ and $A_i=\mathbf{1}^T$. Then, under Assumptions~\ref{ass:1} and~\ref{ass:2}, for every $\alpha\in\R$, it holds that $\pi$ and 
    $\overline{\pi}=\pi+\alpha v$, where $v_k=\frac{1}{S_{kk}}$,
    yield the same Nash equilibrium of $G_0$.
\end{corollary}
\medskip
\noindent The proof of Corollary~\ref{cor:1} is given in Appendix~\ref{app:proofcor1}.
\medskip

\noindent Taking this into account, it is now clear that for some games the unique solution of the leader's problem does not exist. Therefore, we shift our focus to the local Stackelberg equilibria $\left(l\text{-SE}\right)$ and adopt the notion previously analyzed in~\cite{7088583,10.2307/25147124,9424958}. For the sake of completeness, we repeat it in the following definition.}{
Note that the bi-level optimization problem of the leader is in general non-convex. Moreover, there exist subclasses of the Stackelberg pricing games for which there cannot be a unique minimizer of the leader's objective (see the extended version of the paper). Therefore, we shift our focus to the local Stackelberg equilibria $\left(l\text{-SE}\right)$ and adopt the notion previously analyzed in~\cite{7088583,10.2307/25147124,9424958}.
} 
\begin{definition}[Local Stackelberg equilibrium ($l\text{-SE}$) \cite{9424958}]\label{def:lSE}
    Let $G_1$ be a Stackelberg game as in Definition~\ref{def:SG}. A pair $\left(\hat{x}^*,\hat{\pi}\right)\in\text{gph}\left(\mc N_0\right)\cap\left(\mc X\times\mc P\right)$ is a local Stackelberg equilibrium of $G_1$ if there exist open neighborhoods $\Omega_{\hat{x}^*}$, $\Omega_{\hat{\pi}}$ of $\hat{x}^*$ and $\hat{\pi}$ respectively, such that
    \begin{equation}
        \label{eq:lSE}
        J^L\left(\hat{x}^*,\hat{\pi}\right)\leq\inf_{\left(x^*,\pi\right)\in\text{gph}\left(\mc N_0\right)\cap\Omega}J^L\left(x^*,\pi\right)\,,
    \end{equation}
    where $\Omega\defineas\Omega_{\hat{x}^*}\times\left(\mc P\cap\Omega_{\hat{\pi}}\right)$.
\end{definition}
\medskip
Note that since Proposition~\ref{prop:1} guarantees uniqueness of the Nash equilibrium for any $\pi\in\mc P$, we have that $|\mc N_0(\pi)|=1$. This implies that it suffices to find the local optimum of $J^L$ as a function of $\pi$, i.e., to find $\hat{\pi}$ and its neighborhood $\Omega_{\hat{\pi}}$, since the condition~\eqref{eq:lSE} will always be fulfilled for $\hat{x}^*=\mc N_0(\hat{\pi})$ and the open ball of radius $R$ given by   $\Omega_{\hat{x}^*}\defineas\left\{x\in\mc X \mid \norm{x-\hat{x}^*}<R\right\}$, where $R$ is defined by $R>\max_{x\in\hat{\mc N}\left(\Omega_{\hat{\pi}}\right)}\norm{x-\hat{x}^*}$ and $\hat{\mc N}\left(\Omega_{\hat{\pi}}\right)=\bigcup_{\pi\in\Omega_{\hat{\pi}}}\mc N_0\left(\pi\right)$. Using the Implicit function theorem~\cite{Implicit}, we will show that $J^L\left(x^*,\pi\right)$ is continuously differentiable at every $\pi\in\mc P$, allowing us in return to design a gradient-based method to compute the local Stackelberg equilibria.
\section{Algoritm for computing the local Stackelberg equilbirium}\label{sec:alg}
In this section, we show how the local Stackelberg equilibrium given by Definition~\ref{def:lSE} can be computed using a gradient descent method based on the Implicit function theorem.
\subsection{Projected Gradient descent with Armijo rule}
In order to find the $l\text{-SE}$, we adopt a projected gradient descent method with 'Armijo step-size rule along the projection arc'~\cite{Bertsekas/99} to iteratively update $\pi$. Let $\beta$, $\overline{s}$ and $\delta$ be fixed scalars such that $\beta\in\left(0,1\right)$, $\overline{s} >0$ and $\delta\in\left(0,1\right)$. Moreover, let us define the mapping $\pi^{+}:\mc P\times\R_{>0}\rightarrow\mc P$ as
\begin{equation}
\label{eq:pgd1}
    \pi^{+}\left(\pi_t, s\right)\defineas \Pi_{\mc P}\left[\pi_t-\left.s\frac{\text{d} J^L\left(x^*\left(\pi\right),\pi\right)}{\text{d}\pi}\right\vert_{\pi=\pi_t}\right]\,,
\end{equation}
where $\Pi_{\mc P}$ is the projection operator on the leader's constraint set and $x^*\left(\pi\right)$ emphasizes dependence of the Nash equilibrium on $\pi$. The leader's strategy is then updated as
\begin{equation}
\label{eq:pgd2}
    \pi_{t+1}=\pi^{+}(\pi_t, s_t)\,,
\end{equation}
where $s_t=\beta^{l\left(t\right)}\overline{s}$ and $l\left(t\right)\in\Z_{\geq0}$ is the smallest nonnegative integer such that 
\begin{equation}
\label{eq:pgd3}
    \label{eq:termcond}
    \begin{split}
    &J^L\left(\cdot,\pi_t\right)-J^{L}\left(\cdot,\pi^+\left(\pi_t, s_t\right)\right)\geq \\
    & \geq\delta\left(\left.\frac{\text{d} J^L\left(\cdot,\pi\right)}{\text{d}\pi}\right\vert_{\pi=\pi_t}\right)^T\left(\pi_t-\pi^{+}\left(\pi_t, s_t\right)\right)\,.
    \end{split}
\end{equation}
It is important to note that the Armijo step size is well defined, i.e., an appropriate step size will be found after a finite number of trials for $J^L(\cdot, \pi)$ that is continuously differentiable on a specific set $\mc P$~\cite[P2.3.3]{Bertsekas/99}. Hence, we make a standing assumption on the leader's constraint set.
\begin{assumption}\label{ass:3}
    Set $\mc P$ is nonempty, closed and convex.
\end{assumption}
\medskip
The complexity of each update step now boils down to ensuring that $J^L\left(x^*\left(\pi\right),\pi\right)$ is differentiable at $\pi_t$, i.e., showing that the gradient of the leader's cost with respect to the current strategy given by
\begin{equation}
\label{eq:derivative1}
    \frac{\text{d} J^L\left(\cdot\right)}{\text{d} \pi}=\frac{\partial J^L\left(\cdot\right)}{\partial \pi}+\left(\frac{\text{d} x^*}{\text{d} \pi}\right)^T\frac{\partial J^L\left(\cdot\right)}{\partial x^*}\,,
\end{equation}
is well defined. The difficult part of computing~\eqref{eq:derivative1} stems from the need to compute $\frac{\text{d} x^*}{\text{d} \pi}$, i.e., from having to estimate how the Nash equilibrium of $G_0\left(\pi\right)$ reacts to any change in $\pi$. Since the constraint sets of the followers $\mc X_i$ are decoupled, by transforming the problem~\eqref{eq:derivative1} into a form given by 
\iftoggle{full_version}{
\begin{equation}
\label{eq:derivative2}
    \frac{\text{d} J^L\left(\cdot\right)}{\text{d} \pi}=\frac{\partial J^L\left(\cdot\right)}{\partial \pi}+\sum_{i\in\mc I}\textbf{D}^T_{\pi}x^{i*}\frac{\partial J^L\left(\cdot\right)}{\partial x^{i*}}\,,
\end{equation}
}{
$\frac{\text{d} J^L\left(\cdot\right)}{\text{d} \pi}=\frac{\partial J^L\left(\cdot\right)}{\partial \pi}+\sum_{i\in\mc I}\textbf{D}^T_{\pi}x^{i*}\frac{\partial J^L\left(\cdot\right)}{\partial x^{i*}}$, 
} we can organize the complete system in a semi-decentralized form. Knowing that the Nash equilibrium computed via~\eqref{eq:iterativeproc} has to solve the best-response optimization problems of the followers, we show that the individual Jacobians $\textbf{D}_{\pi}x^{i*}$ are well defined. Taking into account the necessary conditions for the Implicit function theorem to hold, we formulate an optimization problem equivalent to the one given by~\eqref{eq:best_resp} for every $i\in\mc I$, and directly apply the theorem on the problem's KKT mapping $l_i\left(z_i,\pi\left|x^{-i*}\right.\right)$ in order to obtain $\textbf{D}_{\pi}x^{i*}$. 
\subsection{Differentiating the KKT conditions}
Based on~\cite{Implicit}, to locally compute the Jacobians $\textbf{D}_{\pi}x^{i*}$, we aim to apply the Implicit function theorem on the mapping $l_i\left(z_i,\pi\left|x^{-i*}\right.\right)$. For every $i\in\mc I$, let the solution mapping $z^{i*}:\mc P\rightrightarrows \mc X_i\times\R_{\geq0}^{m_i^{\text{ineq}}}\times\R^{m_i^{\text{eq}}}$ be
\begin{equation}
\label{eq:solmap}
    z^{i*}\left(\pi\right)=\left\{z^i\left|l_i\left(z_i,\pi\left|x^{-i*}\right.\right)=0\right.\right\}\,.
\end{equation}  
Moreover, let the set of constraints $\Gamma_i$ be defined as
\iftoggle{full_version}{
\begin{equation}
    \label{eq:set_ass3}
    \Gamma_i\defineas\left\{j\in\bigl[1,m_i^{\text{ineq}}\bigr]\cap\N\mid (\hat{\lambda}_i)_j=0 \land \boldsymbol{g_1^i}(x^i)_j=0\right\}\,,
\end{equation}
}{
$\Gamma_i\defineas\left\{j\in\bigl[1,m_i^{\text{ineq}}\bigr]\cap\N\mid (\hat{\lambda}_i)_j=0 \land \boldsymbol{g_1^i}(x^i)_j=0\right\}$,
} where $\boldsymbol{g_1^i}(x^i)_j$ represents the $j$-th inequality constraint. Then the Implicit function theorem from~\cite{Implicit} adapted to our problem reads as the following theorem.
\begin{theorem}\label{th:1}
    For each agent $i\in\mc I$, let the Assumptions~\ref{ass:1} and~\ref{ass:2} hold and $\left(x^{i*},x^{-i*}\right)$ be the result of the Picard-Banach iteration in Proposition~\ref{prop:1}. Furthermore, let the best-response optimization problem be defined via~\eqref{eq:best_resp}, its KKT mapping $l_i\left(z_i,\pi\left|x^{-i*}\right.\right)$ be defined as~\eqref{eq:KKToperator}, $z^{i*}\left(\pi\right)$ be its solution map according to~\eqref{eq:solmap} and the set of constraints $\Gamma_i$  
    be empty. If $l_i\left(\hat{z}_i,\pi\left|x^{-i*}\right.\right)=0$ and $\textbf{D}_{z_i}l_i\left(\hat{z}_i,\pi\left|x^{-i*}\right.\right)$ is non singular for some $\hat{z}_i$, then the solution mapping has a single-valued localization $z^{i*}$ around $\hat{z}_i=(\hat{x}^i,\hat{\lambda}_i,\hat{\nu}_i)$, that is continuously differentiable in a neighbourhood $\mc O$ of $\pi$ with the Jacobian 
    \begin{equation}
        \label{eq:ImplicitJacobian}
        \textbf{D}_{\pi}z^{i*}\left(\pi\right)=-\textbf{D}_{z_i}^{-1}l_i\left(\hat{z}_i,\pi\left|x^{-i*}\right.\right)\textbf{D}_{\pi}l_i\left(\hat{z}_i,\pi\left|x^{-i*}\right.\right)\,,
    \end{equation}
    where $\textbf{D}_{z_i}l_i\left(\hat{z}_i,\pi\left|x^{-i*}\right.\right)$ and $\textbf{D}_{\pi}l_i\left(\hat{z}_i,\pi\left|x^{-i*}\right.\right)$ satisfy
    \small
    \begin{equation}
        \textbf{D}_{z_i}l_i\defineas\left[\begin{array}{ccc}\textbf{D}_{x^i}\nabla_{x^i}L\left(\hat{z}^i,\pi\right) & \textbf{D}_{x^i}^T\boldsymbol{g_1^{i}}\left(\hat{x}^i\right) & \textbf{D}_{x^i}^T\boldsymbol{g_2^{i}}\left(\hat{x}^i\right) \\ \text{Dg}\left(\hat{\lambda}_i\right)\textbf{D}_{x^i}\boldsymbol{g_1^{i}}\left(\hat{x}^i\right) & \text{Dg}\left(\boldsymbol{g_1^{i}}\left(\hat{x}^i\right)\right) & \textbf{0}\\
        \textbf{D}_{x^i}\boldsymbol{g_2^{i}}\left(\hat{x}^i\right) & \textbf{0} & \textbf{0}\end{array}\right]
    \end{equation}
    \begin{equation}
        \textbf{D}_{\pi}l_i\defineas\left[\begin{array}{c}\textbf{D}_{\pi}\nabla_{x^i}L\left(\hat{z}^i,\pi\right)\\ \text{Dg}\left(\hat{\lambda}_i\right)\textbf{D}_{\pi}\boldsymbol{g_1^{i}}\left(\hat{x}^i\right)\\
        \textbf{D}_{\pi}\boldsymbol{g_2^{i}}\left(\hat{x}^i\right)\end{array}\right]\,.
    \end{equation}
    \normalsize
\end{theorem}
\medskip

\noindent Note that for $\hat{z}_i=(x^{i*},\lambda_i^{*},\nu_i^{*})$, where $x^{i*}$, $\lambda_i^{*}$ and $\nu_i^{*}$ describe the unique solution of~\eqref{eq:best_resp}, it directly holds that $l_i\left(\hat{z}_i,\pi\left|x^{-i*}\right.\right)=0$. However, in order to extract the derivative $\textbf{D}_{\pi}x^{i*}$ from $\textbf{D}_{\pi}z^{i*}\left(\pi\right)$ using the Implicit function theorem, we need to make sure that the matrix $\textbf{D}_{z_i}l_i\left(\hat{z}_i,\pi\left|x^{-i*}\right.\right)$ is invertible and that $\Gamma_i$ is empty. Generally speaking, these two conditions directly depend on two things, i.e., the structure of the game resulting from the nature of the application and the Nash equilibrium computed via the Picard-Banach iteration as in Proposition~\ref{prop:1}. To avoid the potential problems, we leverage the benefits of having the unique Nash equilibrium precomputed before every update step of the leader. In the following section, we propose to differentiate the KKT conditions of an optimization problem equivalent to~\eqref{eq:best_resp} that will guarantee $\textbf{D}_{z_i}l_i\left(\hat{z}_i,\pi\left|x^{-i*}\right.\right)$ is invertible and $\Gamma_i$ is empty.
\subsection{Equivalent best-response optimization problem}
Let for every agent $i\in\mc I$ the output of the Picard-Banach iteration be the point $x^{i*}\in\mc X_i$ and $\mc A\left(x^{i*}\right)$ represent the set of all active inequality constraints at $x^{i*}$
\begin{equation}
\label{eq:ax}
    \mc A\left(x^{i*}\right)\defineas\left\{\left.j\in\left[1,m_i^{\text{ineq}}\right]\cap\N\right|\left(g_i^j\right)^Tx^{i*}=\left(h_i\right)_j\right\} \,,
\end{equation}
where $g_i^j$ represents a vector corresponding to the $j-$th row of $G_i$. Moreover, we define the complement of $\mc A\left(x^{i*}\right)$ as 
\begin{equation}
\label{eq:axplus}
    \mc A^{\dagger}\left(x^{i*}\right)=\left(\left[1,m_i^{\text{ineq}}\right]\cap\N\right)\setminus\mc A\left(x^{i*}\right)\,.
\end{equation}
If $\left|\mc A\left(x^{i*}\right)\right|\neq 0$, we let $\overline{G}_i\in\R^{\left|\mc A\left(x^{i*}\right)\right|\times m_F}$ be a matrix whose rows are the rows of $G_i$ listed in $\mc A\left(x^{i*}\right)$. Similarly, we let $\underline{G}_i\in\R^{\left|\mc A^\dagger\left(x^{i*}\right)\right|\times m_F}$ consist of all the remaining rows of $G_i$ and split the vector $h_i$ into $\overline{h}_i$ and $\underline {h}_i$ such that $\overline{G}_ix^{i*}=\overline{h}_i$ and $\underline{G}_ix^{i*}<\underline{h}_i$. This allows us to define an auxiliary best-response optimization problem equivalent to~\eqref{eq:best_resp}.
\begin{corollary}\label{cor:2}
    For each agent $i\in\mc I$, let $\left(x^{i*},x^{-i*}\right)$ be the result of the Picard-Banach iteration in Proposition~\ref{prop:1}. Moreover, let $\mc A\left(x^{i*}\right)$ and $\mc A^{\dagger}\left(x^{i*}\right)$ be defined as in~\eqref{eq:ax} and~\eqref{eq:axplus}. If $\left|\mc A\left(x^{i*}\right)\right|\neq 0$ then $x^{i*}$, such that $\overline{G}_ix^{i*}=\overline{h}_i$ and $\underline{G}_ix^{i*}<\underline{h}_i$, solves the best-response optimization problem given in~\eqref{eq:best_resp} if and only if it solves the optimization problem
    \begin{equation}
    \overline{G}_0^{i}\left(\pi\right):=\left\{\begin{array}{c}
    \displaystyle\min_{x^i\in\R^{m_F}}J^i\left(x^i,x^{-i*},\pi\right) \\
    \text {s.t.}\left[\begin{array}{c}A_i \\ \overline{G}_i\end{array}\right] x^i=\left[\begin{array}{c}b_i \\ \overline{h}_i\end{array}\right] \land \underline{G}_ix^i\leq \underline{h}_i
    \end{array}\right\} \,.
    \label{eq:best_resp2}
\end{equation}
\end{corollary}
\medskip
Let us denote $\overline{A}_i^T=\left[A_i^T \quad  \overline{G}_i^T\right]$ and $\overline{b}_i^T=\left[b_i\quad \overline{h}_i^T\right]$. Then, under the assumption that the linear equality constraints imposed by $\overline{A}_i$ are independent, we can guarantee that the conditions for Theorem~\ref{th:1} will hold for the equivalent best-response optimization problem defined via~\eqref{eq:best_resp2}. 
\begin{assumption}\label{ass:4}
For every $i\in\mc I$, matrix $\overline{A}_i$ is full row rank.
\end{assumption}
\medskip
\begin{theorem}\label{th:2}
    Let the auxiliary best-response optimization problem be defined as in~\eqref{eq:best_resp2}, $x^{i*}\in\mc X_i$ be the optimal value computed as in Proposition~\ref{prop:1} under the Assumptions~\ref{ass:1} and~\ref{ass:2}. If in addition the Assumption~\ref{ass:4} holds, then the set $\overline{\Gamma}_i$ associated with $\overline{G}_0^{i}\left(\pi\right)$ is empty and the matrix $\textbf{D}_{z_i}l_i\left(\hat{z}_i,\pi\left|x^{-i*}\right.\right)$ associated with $\overline{G}_0^{i}\left(\pi\right)$ and defined in Theorem~\ref{th:1} is invertible and reduces to
    \small
    \begin{equation}
    \label{eq:Kbar}
    \textbf{D}_{z_i}l_i\left(\hat{z}_i,\pi\left|x^{-i*}\right.\right)=\left[\begin{array}{c:cc}
    \nabla_{x^ix^i} J^i & \left(\underline{G}_i\right)^{\top} & \overline{A}_i^{\top} \\
    \hdashline
    \mathbf{0} & \operatorname{Dg}\left(\underline{G}_ix^{i*}-\underline{h}_i\right) & \mathbf{0} \\
    \overline{A}_i & \mathbf{0} & \mathbf{0}
    \end{array}\right]\,,    
    \end{equation}
    \normalsize
\end{theorem}
\iftoggle{full_version}{
\medskip
\noindent The proof of Theorem~\ref{th:2} is given in Appendix~\ref{app:proofth2}.
}{
\medskip
\noindent The proof is given in the extended version of our paper.}
\medskip
\iftoggle{full_version}{
\IncMargin{1em}
\begin{algorithm}
\SetKwData{Left}{left}\SetKwData{This}{this}\SetKwData{Up}{up}
\SetKwFunction{Union}{Union}\SetKwFunction{FindCompress}{FindCompress}
\SetKwInOut{Input}{input}\SetKwInOut{Output}{output}
\Input{$\gamma$, $\beta$, $\overline{s}$, $\delta$,$\varepsilon$, $x_0$, $\pi_0$, $T$}
\Output{Pricing vector $\pi$}
\BlankLine
$x_{-1}=\Pi_{\mc X}\left[x_0+2\varepsilon\textbf{1}\right]$ - random initialization\;
\For {$t\leftarrow 0$ \KwTo $T$}{
$k=0$\;
\While{$\norm{x_{k}-x_{k-1}}>\varepsilon$}{
$\text{Followers }i\in\mc I\text{\:in parallel:}$\
$\:x^i_{k+1}=\Pi_{\mc X_i}\left[x_{k}^i-\gamma\nabla_{x^i}J^i\left(x^i,x^{-i},\pi_t\right)\right]$\;
$k\leftarrow k+1$\;
}
$\text{Followers }i\in\mc I\text{\:in parallel:}$\
$\text{\:Define auxiliary problem\:}\overline{A}_i,\underline{G}_i,\underline{h}_i$\;
$\text{\:\:\:Calculate\:}\textbf{D}_\pi x^{i*}\text{ using}$~\eqref{eq:ImplicitJacobian}\;
$\text{Leader } L:\quad\quad\quad$\
$\frac{\text{d} J^L\left(\cdot\right)}{\text{d} \pi}=\frac{\partial J^L\left(\cdot\right)}{\partial \pi}+\sum_{i\in\mc I}\textbf{D}^T_{\pi}x^{i*}\frac{\partial J^L\left(\cdot\right)}{\partial x^{i*}}$\;
$\:\:\:s_t=\text{ArmijoStep}\left(\beta, \overline{s},\delta,\pi_t,\frac{\text{d} J^L\left(\cdot\right)}{\text{d} \pi}\right)$\;
$\:\:\:\pi_{t+1}=\pi^+\left(\pi_t,s_t\right)$\;
}
\caption{Computing leader's strategy}\label{alg:initdir}
\end{algorithm}\DecMargin{1em}
}

\subsection{Convergence of the algorithm}
\iftoggle{full_version}{
Having described how to calculate the gradient of the leader's objective with respect to $\pi$, we can summarize our iterative procedure for computing the local Stackelberg equilibrium in Algorithm~\ref{alg:initdir}. Moreover, we provide the formal convergence guarantees in the following theorem.}{
Finally, we provide formal convergence guarantees.
}
\begin{theorem}\label{th:3}
    Let the Stackelberg game be defined as in~\eqref{eq:SG} and its auxiliary best-response optimization problem as~\eqref{eq:best_resp2}. Let the  Assumptions~\ref{ass:1},~\ref{ass:2},~\ref{ass:3} and~\ref{ass:4} hold and $\left\{\pi_t\right\}$ be the sequence generated by the projected gradient descent method defined by equations~\eqref{eq:pgd1},~\eqref{eq:pgd2} and~\eqref{eq:pgd3}. Then it holds that
    \begin{equation}
    \label{eq:lim}
        \lim_{t\rightarrow+\infty}\left[J^L\left(\cdot,\pi_{t+1}\right)-J^L\left(\cdot,\pi_t\right)\right]=0\,,
    \end{equation}
    and every limit point of $\left\{\pi_t\right\}$ is stationary.
\end{theorem}
\iftoggle{full_version}{
\medskip
\noindent The proof of Theorem~\ref{th:3} is given in Appendix~\ref{app:proofth3}.
}{
\medskip
\noindent The proof is given in the extended version of our paper.}
\medskip
\section{Numerical Example}\label{sec:example}
\iftoggle{full_version}{
\begin {figure}
\centering
\input{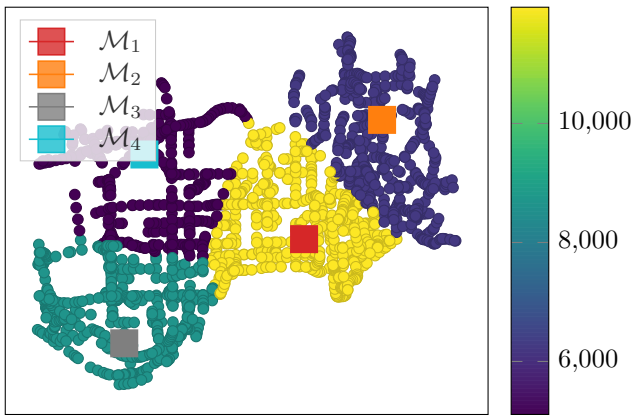}
\caption{Map of Shenzhen - The area divided into 4 regions around charging stations $\mc M=\left\{\mc M_1,\mc M_2,\mc M_3,\mc M_4\right\}$. Color of the nodes within each region indicates the total number of ride-hailing requests whose origin is in that region. The highest number of requests occurs in the region around $\mc M_1$ whereas the lowest number occurs in the region around $\mc M_4$.}
\label{fig:case_study}
\end{figure}}{}
In this section, we provide an illustrative example that tackles the problem of balancing the charging demand of electric ride-hailing fleets on public charging stations. 
We consider a case study introduced in~\cite{Hierarchical} where three ride-hailing companies $\mc C=\left\{\mc C_1, \mc C_2, \mc C_3\right\}$ operate in the Shenzhen area divided into four regions by Voronoi partition around the public charging stations $\mc M=\left\{\mc M_1, \mc M_2, \mc M_3, \mc M_4\right\}$. During a rush-hour period of three hours, the electric vehicles (EVs) serve the heterogeneously distributed requests that appear in the system according to the real taxi demand from~\cite{BEOJONE2021102890}\iftoggle{full_version}{
and shown in Figure~\ref{fig:case_study}.}{.} After the rush-hour period, we assume that EVs with a battery below $60\%$ would go to recharge. 

We assume it is the ride-hailing company's operator who decides how to split the vehicle fleet among different charging stations in an attempt to minimize the cumulative operational cost of the company. Since the companies use the same charging stations, they inherently have to compete for the resources and hence play a game. The model of the company is taken from~\cite{Hierarchical} so the operator's decision vector is given by $x^i\in\R^4$, where $x^i_j\geq 0$ denotes the number of vehicles directed to station $j$ and $\textbf{1}^T_4x^i=N_i$, where $N_i$ is the total number of company's EVs that need to be recharged. The operational cost of each company consists of the charging cost, the cost of queuing at different charging stations due to their limited capacities, and the negative expected revenue in the regions around charging stations. 
\iftoggle{full_version}{
The charging cost is given by $J^i_1\left(x^i,\pi\right)=\left(x^i\right)^TS_i\pi$, where $S_{i} \in \R^{4 \times  4}$ is diagonal, $S_i\succeq 0$ and the entries $\left(S_{i}\right)_{kk}$ can be interpreted as the expected average charging demand per vehicle when choosing the station $k$. The queuing cost and the negative expected revenue are given by $J_{2}^{i}\left(x^{i}, \sigma(x^{-i})\right)=\left(x^{i}\right)^{T}Q\left(x^i+\sigma(x^{-i})-M\right)$ and $J_{3}^{i}\left(x^{i}\right)=\left(e_{i}^{\text{arr}}\right)^{T}x^{i}-\left(e_{i}^{\text{pro}}\right)^{T}x^{i}$, where $Q\in\R^{4\times 4}$ is a positive definite diagonal scaling matrix whose diagonal entries describe how expansive  it is to queue in the regions around charging stations, $M\in\R^{4}$ is the vector of charging stations' capacities, $e_{i}^{\text{arr}}\in \mathbb{R}^{ 4}$ is the average cost of a vehicle being unoccupied while traveling to a station and the vector $e_{i}^{\text{pro}}\in \mathbb{R}^{4}$ is the expected profit in regions around charging stations estimated from historical data. }
{}The linear inequality constraints encompass information about the number of vehicles that can reach a certain station based on a linear battery discharge model and current battery level after the three-hour simulation. For a detailed description of all parameters, we refer the reader to~\cite{Hierarchical}. 

The competition between the companies now matches the structure of the game $G_0\left(\pi\right)$, where $\pi\in\R_+^4$ is the vector of charging prices at different stations that the central authority of the region sets. This central authority, i.e, the leader, can be anything from a power-providing company interested in balancing the demand on the power grid to the government trying to design the pricing incentives that would increase the coverage and motivate the idle taxi drivers not to flock in the more demand attractive areas. Therefore, we assume that the central authority defines a set point vector $N_{\text{des}}\in\R^4$ describing the desired number of vehicles at each charging station and plays the game $G_1$ with the ride-hailing companies in an attempt to minimize the cost $J^L\left(x^*\left(\pi\right),\pi\right)=\frac{1}{2}\norm{\sigma\left(x^*\left(\pi\right)\right)-N_{\text{des}}}^2_2$ 
by choosing adequate $\pi\in\mc P$. We set $\mc P\defineas[p_{\text{min}},p_{\text{max}}]^{4}$, $p_{\text{min}}=1.0$ and $p_{\text{max}}=5.0$. The central authority's set point vector is chosen as $N_{\text{des}}=\left[198, 103, 144, 87\right]$ to match the demand distribution among the regions and the other parameters are given by $\beta=0.25$, $\overline{s}=0.000001$ and $\delta=0.00001$.

In~\cite{Hierarchical}, an exhaustive grid search procedure over the price domain $\mc P$ was used to find an approximation of the global minimum of $J^L\left(x^*\left(\pi\right),\pi\right)$. After 25000 iterations, the two best pricing vectors $\overline{\pi}_1=\left[2.75, 1.625, 2.208, 1.0\right]$ and $\overline{\pi}_2=\left[4.03, 2.8, 3.49, 2.24\right]$, that yield $J^L\left(x^*\left(\overline{\pi}_1\right),\overline{\pi}_1\right)=13.585$ and $J^L\left(x^*\left(\overline{\pi}_2\right),\overline{\pi}_2\right)=18.579$, were found. 
\begin{figure}
    \centering
\input{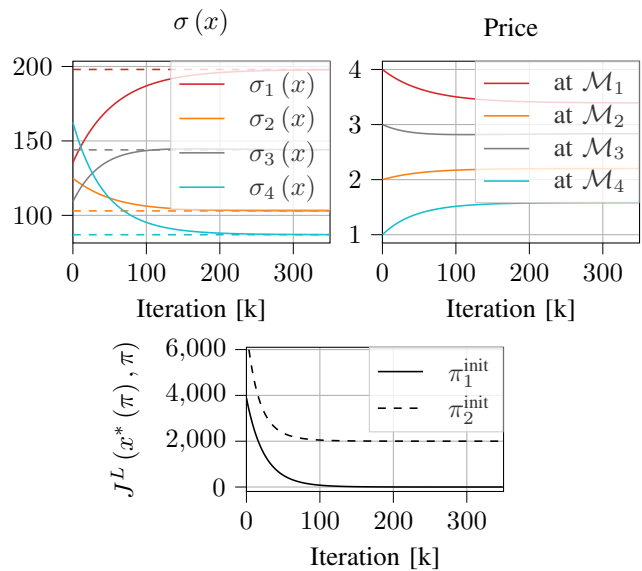}
    \caption{The upper plots show the evolution of the total vehicle accumulation at the charging stations $\sigma(x)$ and the charging prices $\pi$ during the iterative procedure and for the initial charging price $\pi^{\text{init}}_1$. The lower plot compares the evolution of the leader's objective $J^L\left(x^*\left(\pi\right),\pi\right)$ when starting from different initial prices, i.e., $\pi^{\text{init}}_1$ and $\pi^{\text{init}}_2$.\vspace{-0.5cm}}
    \label{fig:accprice}
\end{figure}
In Figure~\ref{fig:accprice}, we show how the gradient-based iterative procedure can be used to find the local Stackelberg equilibrium as per Definition~\ref{def:lSE}. The upper left plot in the figure shows the distribution of ride-hailing fleets among different charging stations, whereas the upper right plot shows how the prices of charging evolve with respect to the iteration number when the initial value of the leader's decision variable is $\pi^{\text{init}}_1=\left[4.0,2.0,3.0,1.0\right]$. The resulting vector of charging prices found by the proposed method is given by  $\pi^*=\left[3.394,2.201,2.833,1.584\right]$ and yields $J^L\left(x^*\left(\pi^*\right),\pi^*\right)=2.2\cdot 10^{-5}$. Clearly, the attained value of the leader's objective is significantly smaller than the ones obtained using only the grid-search procedure. 

In the lower plot, however, we show that the attained $l-$SE largely depends on the initial value of the leader's decision variable. Namely, even though the algorithm was able to find the $l-$SE corresponding to the global minimum of the leader's objective when starting from $\pi^{\text{init}}_1$, it is also the case that starting from $\pi^{\text{init}}_2=\left[3.0,3.0,3.0,3.0\right]$ yields an equilibrium with a significantly higher value of the leader's objective. Given the local-search nature of the proposed procedure, in order to find the $l-$SE corresponding to the global minimum of the leader's objective, we would have to combine this iterative method with a higher-level heuristic.  
\section{Conclusions}\label{sec:conclusion}
In this paper, we present an iterative, semi-decentralized, gradient-based method for finding the $l-$SE for a particular class of Stackelberg games. Essentially, if every follower locally differentiates the KKT conditions of an optimization problem equivalent to its best-response optimization problem, a framework that guarantees convergence to a local Stackelberg equilibrium can be established. The attained equilibrium, however, depends largely on the starting point of the iterative procedure, and developing intelligent higher-level heuristics will be a future research direction.    

\bibliographystyle{IEEEtran}
\bibliography{references.bib}

\iftoggle{full_version}{
\appendix
\subsection{Proof of Proposition~\ref{prop:1}}\label{app:proof1}
\begin{proof}
    Since $P_i\succ 0$ for every $i\in\mc I$ under the Assumption~\ref{ass:1}, the agents' cost functions are convex in $x^i$. Under the Assumption~\ref{ass:2}, based on~\cite[T.1]{Rosen}, there exists a Nash equilibrium of $G_0\left(\pi\right)$ for any $\pi\in\mc P$. According to~\cite[Ch.2]{VIproblems}, a sufficient condition for the uniqueness of the Nash equilibrium is that the operator $F\left(x,\pi\right)$ is strictly monotone in $x$. Based on the structure of the cost function, the pseudo gradient can be written as an affine mapping $F\left(x,\pi\right)=F_1x+F_2$, such that
    $F_1=\mathbb{I}_{N}\otimes\left(P-Q\right)+\mathbf{1}_{N}\mathbf{1}_{N}^T\otimes Q$ and 
    $F_2=\left[r_i+S_i\pi\right]_{i\in\mc I}$. To show that $F\left(x,\pi\right)$ is strictly monotone, it suffices to prove that $F_1\succ 0$~\cite{ConvexAnalysis}. This is true as under the Assumption~\ref{ass:1}, for any $x\in\mc X$, it holds that
    $x^TF_1x=\sum_{i\in\mc I}\left(x^i\right)^T\left(P-Q\right)x^i+\left(\sum_{i\in\mc I}x^i\right)^TQ\left(\sum_{i\in\mc I}x^i\right)>0$. Since the projection operator is non-expansive, for $\gamma$ such that the mapping $x\rightarrow x-\gamma F(x,\pi)$ is contractive, the mapping $x\rightarrow \Pi_{\mc X}\left[x-\gamma F\left(x,\pi\right)\right]$ is also contractive. Hence, the proof that Picard-Banach iteration converges to the Nash equilibrium can be found in~\cite{Paccagnan2016a,Decentralized}. 
\end{proof}
\subsection{Proof of Corollary~\ref{cor:1}}\label{app:proofcor1}
\begin{proof}
    Under the Assumptions~\ref{ass:1} and~\ref{ass:2}, for every $i\in\mc I$, the system of equations~\eqref{eq:KKToperator} representing the KKT conditions of the best-response optimization problem reduces to
\small
\begin{equation}
\label{eq:KKTsimplified}
\left\{\begin{array}{l}
P_ix^{i*}+Q_i\sigma\left(x^{-i*}\right)+r_i+S_i\pi+G_i^{\top} \lambda_i^*+A_i^{\top} \nu_i^*=\mathbf{0} \\
\operatorname{Dg}\left(\lambda_i^*\right)\left(G_i x^{i*}-h_i\right)=\mathbf{0} \\
A_i x^{i*}=b_i
\end{array}\right..
\end{equation}
\normalsize
If for every $i\in\mc I$, it holds that $S_i=S$, then we have that $S_i\overline{\pi}=S_i\pi+\alpha\mathbf{1}$. It is now clear that if the tuple $\left(x^{i*},\lambda_i^*,\nu_i^*,\pi\right)$ solves the system~\eqref{eq:KKTsimplified}, so does the tuple $\left(x^{i*},\lambda_i^*,\nu_i^*-\alpha,\overline{\pi}\right)$ since $A_i^T=\mathbf{1}$. Since $G_0$ has a unique Nash equilibrium for any given $\pi$, this means that both $\pi$ and $\overline{\pi}$ yield the same Nash equilibrium of $G_0$.  
\end{proof}
\subsection{Proof of Theorem~\ref{th:2}}\label{app:proofth2}
\begin{proof}
    We start by noting that since $\underline{G}_ix^{i*}<\underline{h}_i$, we have that $\lambda_i^*=\mathbf{0}$ which reduces $\textbf{D}_{z_i}l_i\left(\hat{z}_i,\pi\left|x^{-i*}\right.\right)$ to~\eqref{eq:Kbar} and sets $\overline{\Gamma}_i=\emptyset$. In order to prove Theorem~\ref{th:2}, we recall the Matrix Inversion Lemma~\cite{Matrix}. Let $M_1\in\R^{p\times p}$ be invertible, $M_2,M_3^T\in\R^{p\times q}$, $M_4\in\R^{q\times q}$ and $M\in\R^{(p+q)\times(p+q)}$, $\text{Sh}\left(M_1\right)\in \R^{q\times q}$ be block matrices $$M\defineas\left[\begin{array}{cc}M_1 & M_2\\M_3 & M_4\end{array}\right],\quad \text{Sh}\left(M_1\right)\defineas\left[\begin{array}{cc}S_1 & S_2 \\ S_3 & S_4\end{array}\right],$$ where $\text{Sh}\left(M_1\right)$ is the Shur complement of $M_1$ in $M$, i.e., $\text{Sh}\left(M_1\right)=M_4-M_3M_1^{-1}M_2$. If $M_1$ is nonsingular, then $M$ is invertible if and only if Shur complement of $M_1$ in $M$ is nonsingular. We can partition $\textbf{D}_{z_i}l_i\left(\hat{z}_i,\pi\left|x^{-i*}\right.\right)$ such that $M_1$, $M_2$, $M_3$ and $M_4$ are given by $M_1=\nabla_{x^ix^i} J^i$,
\begin{equation}
    \label{eq:part23}
    M_2=\left[\begin{array}{cc}\underline{G}_i^{\top} & \overline{A}_i^{\top}\end{array}\right],\quad M_3=\left[\begin{array}{c}\mathbf{0}\\ \overline{A}_i \end{array}\right]\,,
\end{equation}
\begin{equation}
    \label{eq:part4}
    M_4=\left[\begin{array}{cc}\operatorname{Dg}\left(\underline{G}_ix^{i*}-\underline{h}_i\right) & \mathbf{0} \\
    \mathbf{0} & \mathbf{0}\end{array}\right]\,.
\end{equation}
Assumption~\ref{ass:1} ensures that matrix $M_1$ is nonsingular since $\nabla_{x^ix^i} J^i=P_i\succ 0$. The individual blocks of $\text{Sh}\left(M_1\right)$ are $S_1=\operatorname{Dg}\left(\underline{G}_ix^{i*}-\underline{h}_i\right)$, 
$S_2=\mathbf{0}$,
$S_3=-\overline{A}_iP_i^{-1}\left(\underline{G}_i\right)^T$ and 
$S_4=-\overline{A}_iP_i^{-1}\overline{A}_i^T$. Because $\mc A^{\dagger}$ encompasses the inactive inequality constraints, we have that $\operatorname{Dg}\left(\underline{G}_ix^{i*}-\underline{h}_i\right)\prec 0$. Moreover, since $P^{-1}\succ 0$ and matrix $\overline{A}_i$ is full row rank according to Assumption~\ref{ass:4}, we have that $S_4\prec 0$. Finally, because both $S_1$ and $S_4$ are non-singular, we have that $\text{Sh}\left(M_1\right)$, and hence $\textbf{D}_{z_i}l_i\left(\hat{z}_i,\pi\left|x^{-i*}\right.\right)$, are invertible.
\end{proof}
\subsection{Proof of Theorem~\ref{th:3}}\label{app:proofth3}
\begin{proof}
First, note that based on the Armijo rule, the sequence $\left\{J^L\left(\cdot,\pi_t\right)\right\}_{t=1}^{\infty}$ is monotonically nonincreasing. Because $J^L\left(\cdot,\pi\right)$ is continuous in $z^T=[x^*(\pi)^T,\pi]$ and $\mc X\times\mc P$ is compact and convex, there exists $J^L_{\text{min}}\in\R$ such that $J^L\left(\cdot,\pi_t\right)\geq J^L_{\text{min}}$ for all $z\in\mc X\times\mc P$. Since the sequence $\left\{J^L\left(\cdot,\pi_t\right)\right\}_{t=1}^{\infty}$ is now monotonically nonincreasing and bounded, it converges to a finite value implying $\lim_{t\rightarrow +\infty}\left[J^L\left(\cdot,\pi_{t+1}\right)-J^L\left(\cdot,\pi_t\right)\right]=0$. Since Theorem~\eqref{th:2} guarantees that $J^L\left(\cdot,\pi\right)$ is continuously differentiable at every $\pi\in\mc P$, the proof that every limit point of $\left\{\pi_t\right\}$ is stationary follows directly from~\cite[P2.3.3]{Bertsekas/99}.  
\end{proof}
}{}
\end{document}